\documentclass[twocolumn,aps,prl,superscriptaddress]{revtex4-2}
\usepackage[utf8]{inputenc}
\usepackage{amsmath}
\usepackage{bm}
\usepackage{xcolor}
\usepackage{graphicx}
\usepackage{outlines}

\begin{document}

\title{Gate-tunable antiferromagnetic Chern insulator in twisted bilayer transition metal dichalcogenides}

\author{Xiaoyu Liu}
\affiliation{Department of Materials Science and Engineering, University of Washington, Seattle, WA 98195, USA}
\author{Chong Wang}
\affiliation{Department of Materials Science and Engineering, University of Washington, Seattle, WA 98195, USA}
\author{Xiao-Wei Zhang}
\affiliation{Department of Materials Science and Engineering, University of Washington, Seattle, WA 98195, USA}
\author{Ting Cao}
\affiliation{Department of Materials Science and Engineering, University of Washington, Seattle, WA 98195, USA}
\author{Di Xiao}
\affiliation{Department of Materials Science and Engineering, University of Washington, Seattle, WA 98195, USA}
\affiliation{Department of Physics, University of Washington, Seattle, WA 98195, USA}

\begin{abstract}
A series of recent experimental works on twisted MoTe$_2$ homobilayers have unveiled an abundance of exotic states in this system. Valley-polarized quantum anomalous Hall states have been identified at hole doping of $\nu = -1$, and the fractional quantum anomalous Hall effect is observed at $\nu = -2/3$ and $\nu = -3/5$. In this work, we investigate the electronic properties of AA-stacked twisted bilayer MoTe$_2$ at $\nu=-2$ by $k$-space Hartree-Fock calculations. We find that the phase diagram is qualitatively similar to the phase diagram of a Kane-Mele-Hubbard with staggered onsite potential. A noteworthy phase within the diagram is the antiferromagnetic Chern insulator, stabilized by the external electric field. We attribute the existence of this Chern insulator to an antiferromagnetic instability at a topological phase transition between the quantum spin hall phase and a band insulator phase. We highlight that the antiferromagnetic Chern insulator phase is most evident at a twist angle of approximately $4^\circ$. Our research proposes the potential of realizing a Chern insulator beyond $\nu=-1$, and contributes fresh perspectives on the interplay between band topology and electron-electron correlations in moir\'e superlattices.
\end{abstract}

\maketitle

\textit{Introduction.}---
Since the experimental discovery of correlated states and superconductivity in twisted bilayer graphene \cite{cao2018unconventional,cao2018correlated}, two-dimensional moir\'e superlattices have emerged as a revolutionary platform in the study of electron-electron correlations. Apart from graphene moir\'e superlattices, transition metal dichalcogenides (TMD) moir\'e superlattices have been under intensive investigation due to the reduced number of degrees of freedom \cite{wu2018hubbard,wu2019topological}. For examples, heterobilayer TMD moir\'e superlattices have been found to host magnetic phases \cite{regan2020mott,tang2020simulation,zhao2023gate,li2021continuous}, charge ordered phases \cite{regan2020mott,xu2020correlated,huang2021correlated,li2021imaging} and quantum anomalous Hall states \cite{li2021quantum,tao2022valley,zhao2022realization,zhang2021spin,devakul2021magic,pan2022topological,xie2022valley}. Recently, there has been a surge in research focused on homobilayer TMD moir\'e superlattices \cite{pan2020band,angeli2021gamma,xian2021realization,anderson2023programming,cai2023signatures,foutty2023mapping}. Compared with heterobilayers, homobilayer TMD moir\'e superlattices host intrinsic topological band structures \cite{wu2019topological,yu2020giant}. Moreover, both layers' electrons actively contribute to the low-energy physics for homobilayers, which provides a unique opportunity to tune the electronic properties with an out-of-plane electric field \cite{haavisto2022topological,anderson2023programming}.

Recently, a series of works on twisted MoTe$_2$ homobilayers (tMoTe$_2$) have unveiled an abundance of topological states in this system. At hole doping $\nu = -1$ (one hole per moir\'e unit cell), valley-polarized quantum anomalous Hall states are observed \cite{cai2023signatures,zeng2023integer,park2023observation}. At $\nu = -2/3$ and $\nu = -3/5$, fractional Chern insulator is observed at zero magnetic field \cite{cai2023signatures,zeng2023integer,park2023observation}. The emergence of these states arise from quenched kinetic energy in the flat bands \cite{li2021spontaneous,wang2023fractional,reddy2023fractional}. Following the same logic, it would appear that there might not be as many interesting observations at $\nu = -2$, for which the flat bands are completely filled. In this scenario, for electron-electron interactions to play a significant role in low-energy dynamics, the interaction energy has to overcome the band gap. Fortunately, the band gap can be substantially reduced by an external electric field, hinting at the potential for gate-tunable correlated states in twisted tMoTe$_2$. Such prospects add a captivating dimension to the exploration of this fascinating material.

In this work, we investigate the electronic properties of AA-stacked tMoTe$_2$ at $\nu=-2$ by $k$-space Hartree-Fock calculations. The phase diagram of this system reveals four distinct states: an antiferromagnetic Chern insulator, an in-plane antiferromagnetic phase, a quantum spin hall (QSH) phase, and a trivial band insulator (BI). We find that the phase diagram is qualitatively similar to the phase diagram of a Kane-Mele-Hubbard with staggered onsite potential \cite{jiang2018antiferromagnetic}. Within the framework of the Kane-Mele-Hubbard model, we explain the existence of the in-plane antiferromagnetic phase by a spin model. The emergence of the antiferromagnetic Chern insulator is attributed to an instability of the phase boundary between the QSH and the BI phases with respect to an antiferromagnetic perturbation under a sufficiently large Coulomb interaction. Additionally, we investigate the tMoTe$_2$ of different twist angles and find that the antiferromagnetic Chern insulator phase is most prominently observed at twist angle $\sim 4^\circ$. Our work points out the possibility of realizing Chern insulator beyond $\nu=-1$ and provides new insights into the interplay between the band topology and electron-electron correlations in moir\'e superlattices. 

\begin{figure}
\centering
\includegraphics[width=\columnwidth]{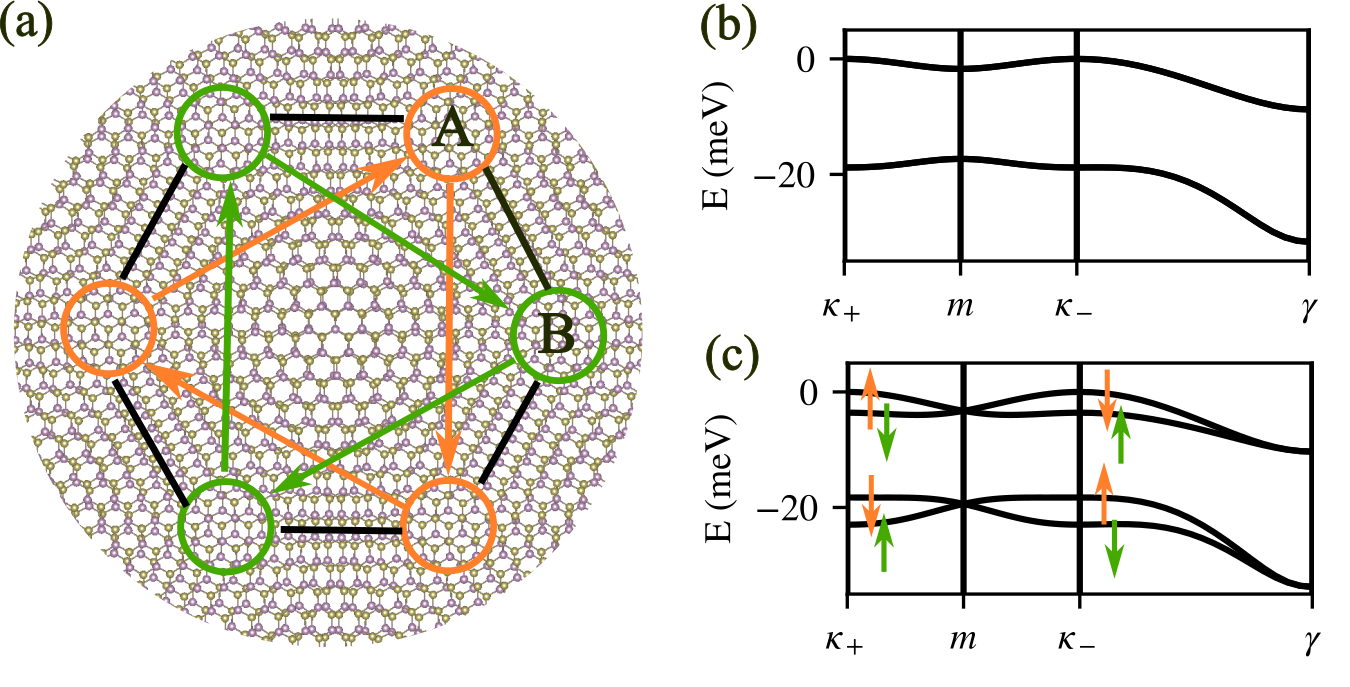}
\caption{(a) The moir\'e superlattice of twisted bilayer tMoTe$_2$. Orange (green) circles represent MX (XM) stackings, where the Mo (Te) atoms sit on top of the Te (Mo) atoms. For the top four moir\'e valence bands, the charge density at MX (XM) stacking is mainly located on bottom (top) layer. In the Kane-Mele-Hubbard model, $\nu_{ij} = 1$ ($\nu_{ij} = -1$) if the hopping is along (opposite to) the direction of the arrows for the next-nearest-neighbor hopping (see main text). (b) The moir\'e band structure without out-of-plane electric field in the mBZ. (c) The moir\'e band structure with an out-of-plane electric field of $\Delta_D = 5~\rm{meV}$. The orange (green) arrows indicate the states are mainly from the bottom (top) layer, and the up (down) arrows indicate spin up (down) states.\label{fig:fig1}}
\end{figure}

\textit{Method.}---
In monolayer MoTe$_2$, the conduction and valence band edges are located at the corners of the Brillouin zone, i.e., $K$ and $K'$ points. Due to strong spin-orbit interaction, electrons in the $K$ and $K'$ valleys have opposite spins \cite{xiao2012coupled}. The major effect of the moir\'e superlattice is to induce coupling between Bloch functions separated by moir\'e reciprocal lattice vectors, which is captured by a continuum model. The separation between the $K$ and $K'$ points are much larger than the length scale of the moir\'e Brillouin zone (mBZ), such that the continuum model is block-diagonal in the valley indices. For $K$ valley, the continuum model Hamiltonian reads \cite{wu2019topological,yu2020giant}:
\begin{equation}\label{eq:continuum_ham}
\begin{aligned}
H_K = & \left(
    \begin{matrix}
        -\frac{\hbar^2(\bm{k}-\bm{\kappa}_+)^2}{2m^*} + \Delta_b(\bm{r})  & \Delta_T(\bm{r}) \\
        \Delta_T^{\dagger}(\bm{r}) & -\frac{\hbar^2(\bm{k}-\bm{\kappa}_-)^2}{2m^*} + \Delta_t(\bm{r})
    \end{matrix}
\right) \\ 
& + \frac{1}{2}\left(\begin{matrix}
    \Delta_D & 0 \\
        0 & -\Delta_D
    \end{matrix}\right),
\end{aligned}
\end{equation}
where the intralayer and interlayer moir\'e potentials are $\Delta_{b/t}=2V\sum_{i=1,3,5}\cos(\bm{G}_i\cdot\bm{r}\pm\phi)$ and $\Delta_T = w(1+e^{-i \bm{G}_2 \cdot \bm{r}} + e^{-i \bm{G}_3 \cdot \bm{r}})$, respectively. $\bm{G}_i = \frac{4\pi}{\sqrt{3} a_M }(\cos\frac{i-1}{3}\pi, \sin\frac{i-1}{3}\pi)$ are moir\'e reciprocal lattice vectors with $a_M$ being the moir\'e lattice constant. $\bm{\kappa}_+ = 2\bm{G}_1 / 3 - \bm{G}_2 / 3$ and $\bm{\kappa}_- = \bm{G}_1 / 3 + \bm{G}_2 / 3$ are the mBZ corners. $m^*$ is the effective mass and is taken as $0.6 m_e$, where $m_e$ is the free electron mass. Layer-differentiating potential proportional to $\Delta_D$ is included in $H_K$ to take into account the out-of-plane electric field. For $K'$ valley, the continuum model Hamiltonian can be deduced by acting time reversal operator on $H_K$.

%This model was found to be a realization of Kane-Mele model \cite{kane2005quantum,kane2005z} with valley-contrasting Chern numbers \cite{wu2019topological,yu2020giant}. 

The parameters for the continuum model is fitted from large-scale density functional theory calculations \cite{wang2023fractional}. Specifically, $(V, \phi, w) = (20.8~\text{meV}, -107.7^\circ,-23.8~\text{meV})$. The corresponding single-particle band structure at twist angle $\theta = 3.89^\circ$ is presented in Fig.~\ref{fig:fig1}(b), where every band is doubly degenerate due to emergent inversion symmetry. The degeneracy can be split by a finite electric field, as shown in Fig.~\ref{fig:fig1}(c).  The topmost moir\'e band from $K$ ($K'$) valley has Chern number $-1$ ($+1$).  Therefore at $\nu = -2$, the Hamiltonian Eq.~(\ref{eq:continuum_ham}) describes a quantum spin Hall insulator~\cite{wu2019topological,yu2020giant}. 

To investigate the effect of electron-electron interaction, we carry out self-consistent Hartree-Fock calculations based on the continuum model. The solution of the continuum model is the envelope function of the atomistic wave function, and the envelope function is expanded as superposition of plane waves. In the basis of the plane waves (labeled by momentum $\bm{k}$), the electron-electron interaction reads
\begin{equation}
H_{\rm int} = \frac{1}{2A} \sum_{l, l', \tau, \tau', \bm{k}, \bm{k}', \bm{q}} V_{ll'} (\bm{q}) {c_{l \tau \bm{k}+\bm{q}}^{\dagger}}  {c_{l' \tau' \bm{k}' -\bm{q}}^{\dagger}}  {c_{l' \tau' \bm{k}'}} c_{l \tau\bm{k}},
\end{equation}
where $A$ is the area of the system, $l$ and $l'$ label layers and $\tau$ and $\tau'$ label valleys. The Coulomb interaction takes the form \cite{chatterjee2020symmetry}
\begin{equation}\label{eq:coulomb}
    V_{ll'}(\bm{q}) = \frac{e^2}{2\epsilon\epsilon_0 |\bm{q}|} \left[\tanh(d_{\rm gate}|\bm{q}|) + (1-\delta_{ll'})(e^{-d |\bm{q}|}-1)\right],
\end{equation}
where $\epsilon$ is the relative dielectric constant, $\epsilon_0$ is the vacuum permittivity, $d_{\rm gate}$ is the distance between the sample and the symmetric metal gate, and $d$ is the distance between the two monolayers. The interlayer Coulomb interaction is reduced from the intralayer Coulomb interaction by $e^{-d |\bm{q}|} - 1$. This correction is only valid for $d \ll d_{\rm gate}$ \cite{chatterjee2020symmetry}. The self-consistent Hartree-Fock calculations are performed in the reciprocal space without projecting the interaction to bands near the charge neutrality point.

\begin{figure}
\centering
\includegraphics[width=\columnwidth]{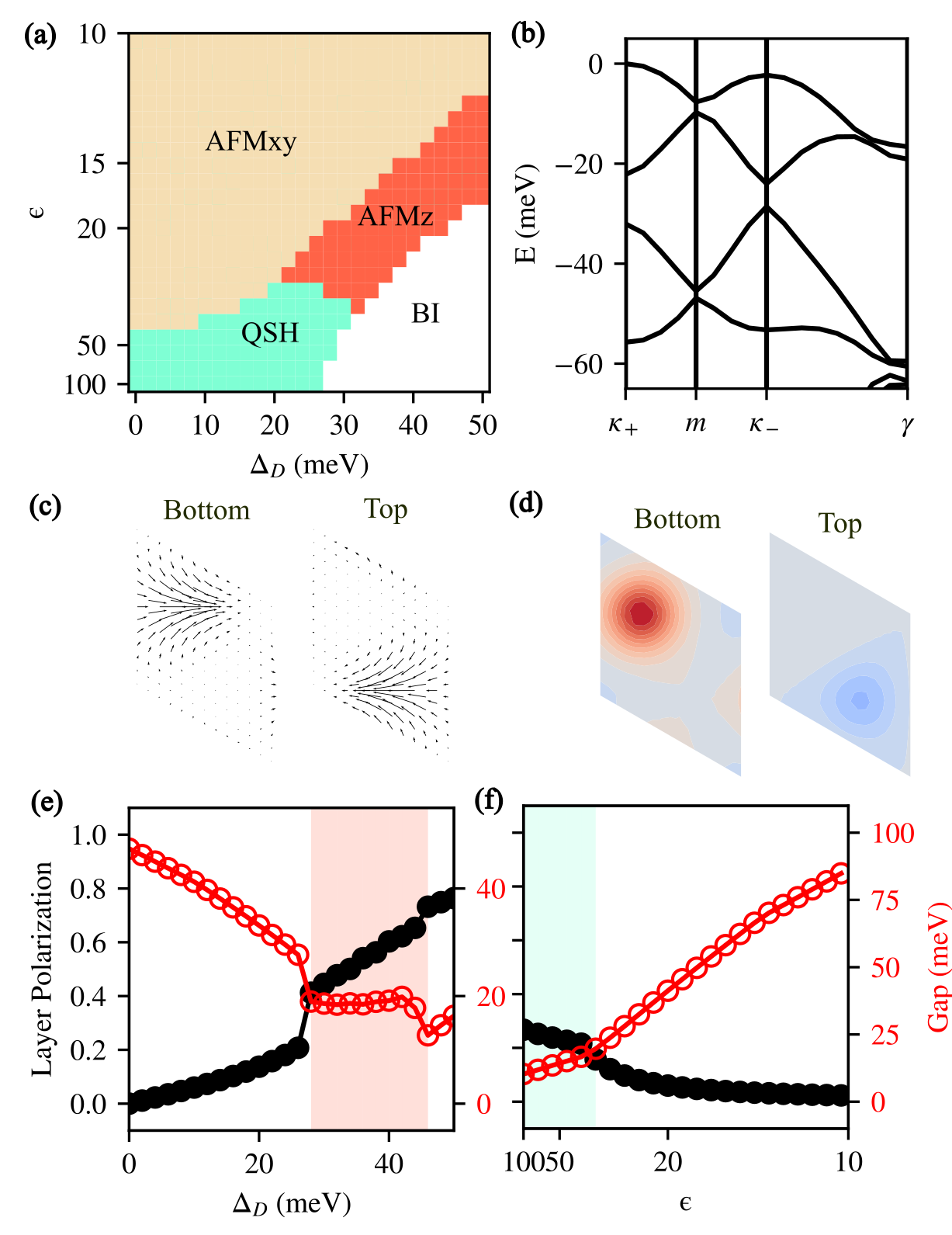}
\caption{(a) Phase diagram of $k$-space Hartree Fock calculation vs displacement field $\Delta_D$ and dielectric constant $\epsilon$. In the calculations, an $18 \times 18$ mesh in the reciprocal space is used and 196 bands are included. (b) The quasiparticle band structure of the AFMz Chern insulator phase at $\Delta_D = 30$~meV and $\epsilon = 33$. (c) In-plane spin texture of the AFMxy phase in the moir\'e unit cell. (d) Out-of-plane spin density of the AFMz phase for both layers. (e) Line cut at $\epsilon=20$. The black dots denote the layer polarization, indicating the difference in hole occupation between the two layers. The red circles denote the charge gap. The red background indicates Chern number equaling 1. Numerical evidences of first order phase transitions can be observed (sudden jump in both charge gap and layer polarization). (f) Line cut at $\Delta_D=10~\rm{meV}$. The aquamarine background indicates the QSH phase. The twist angle for the calculation is 3.89$^\circ$.\label{fig2}}.
\end{figure}

\textit{Phase diagram.}---
Figure \ref{fig2}(a) presents the Hartree-Fock phase diagram as a function of $\epsilon$ and $\Delta_D$ at twist angle $\theta=3.89^\circ$. For noninteracting limit ($\epsilon \to \infty$), the system exhibits a QSH phase at $\Delta_D = 0$. The QSH state can be turned into a trivial BI state via band inversions at $\bm{\kappa}_+$ and $\bm{\kappa}_-$ point for large $\Delta_D$. As the interacting strength increases, two more phases emerge. With $\Delta_D=0$, the AFMxy phase is stabilized for small $\epsilon$. In this phase, the two sublattices possess opposite magnetic moments in the $x-y$ plane [Fig.~\ref{fig2}(c)]. For larger $\Delta_D$, the orientation of the magnetic moment changes to the $\pm z$ direction and the AFMz phase is stabilized [Fig.~\ref{fig2}(d)]. We have also performed Hartree-Fock calculations in a $\sqrt{3}\times\sqrt{3}$ supercell, and no translational symmetry breaking phases are found. The phase transitions from AFMxy to AFMz and from AFMz to BI are characterized as first-order phase transitions [Fig.~\ref{fig2}(e)], whereas the transition from QSH to BI is a second-order phase transition. Based on our numerical evidence, the transition between QSH and AFMxy is more likely to be a second-order phase transition, as shown in Fig.~\ref{fig2}(f).

Most interestingly, the AFMz phase is an antiferromagnetic Chern insulator. The quasiparticle band structure of AFMz features two massive Dirac points [Fig.~\ref{fig2}(b)] at $\bm{\kappa_+}$ and $\bm{\kappa_-}$, both of which contribute $\pi$ Berry flux and the total Chern number is 1. 

\begin{figure}
\centering
\includegraphics[width=\columnwidth]{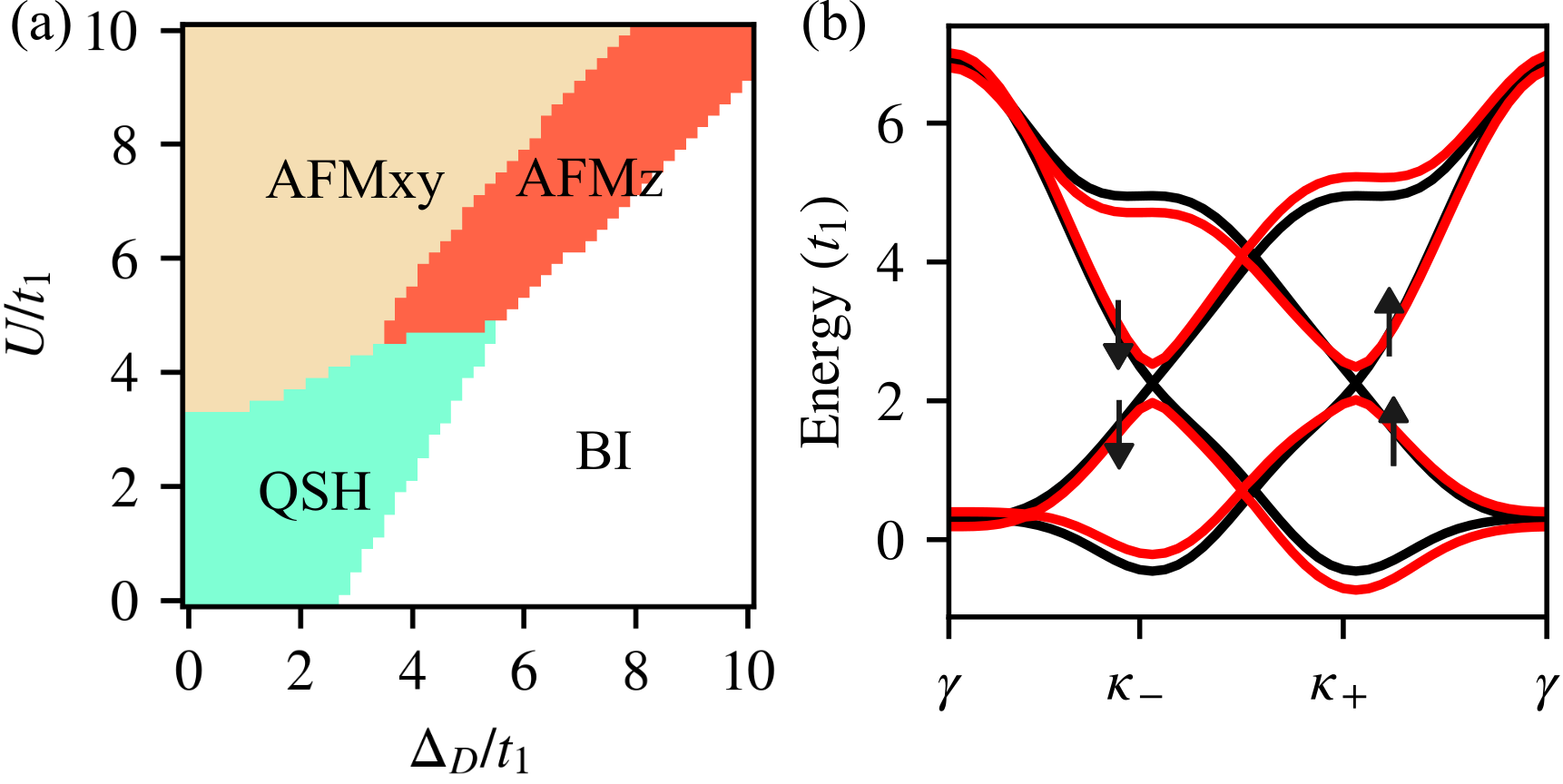}
\caption{(a) The phase diagram of the Kane-Mele-Hubbard model as a function of the Hubbard interaction $U$ and sublattice potential $\Delta_D$. (b) The Hartree-Fock quasiparticle band structure (spins labelled) of the nonmagnetic phase (black) and AFMz (red) phase at $U = 5.4 t_1$ and $\Delta_D = 5.95 t_1$. At this point, the Hartree-Fock ground state is AFMz, and the nonmagnetic phase is obtained by using a nonmagnetic initial state in the self-consistent loop. The nonmagnetic phase is at the transition point between the QSH and NI phases. While the two band structures differ at almost all $\bm{k}$ points, the contribution to the difference of the two total quasiparticle energies mainly comes from the Dirac points at the $K$ and $K'$ points (see main text). The next-nearest-hopping for both (a) and (b) is $t_2 = 0.3 t_1$ and $\theta = \pi/3$.\label{fig:KMH}}
\end{figure}

A real space picture can help understand the phase diagram in Fig.~\ref{fig2}(a). The moir\'e potential has local minimums at the MX (XM) stacking for the bottom and top layers, respectively. As shown in Fig.~\ref{fig:fig1}(a), these local minimums form a buckled honeycomb lattice, where the $A$ ($B$) sublattices are from the bottom (top) layer. Therefore, tMoTe$_2$ can be qualitatively understood as a Kane-Mele-Hubbard model \cite{kane2005quantum,kane2005z,hohenadler2011correlation,hohenadler2012quantum}:
\begin{equation}\label{eq:kmh}
\begin{aligned}
    H_{\rm KMH} = & H_{\rm single} + \sum_i U n_{i\uparrow} n_{i\downarrow}, \\
    H_{\rm single} = & \sum_{\langle i, j\rangle, \sigma} t_1 c_{i\sigma}^\dagger c_{j\sigma} + 
    \sum_{\langle\langle i, j \rangle\rangle, \sigma} t_2 ~ e^{i\sigma\nu_{ij}\theta} c_{i\sigma}^\dagger c_{j\sigma} \\
    & - \frac{1}{2}\sum_{i\in A} \Delta_D c_{i\sigma}^\dagger c_{i\sigma} + \frac{1}{2}\sum_{i\in B} \Delta_D c_{i\sigma}^\dagger c_{i\sigma},
\end{aligned}
\end{equation}
where $t_1$ is the nearest-neighbor hopping, $t_2$ is the next-nearest-neighbor hopping amplitude, and $U$ is the onsite Coulomb interaction. In Eq.~(\ref{eq:kmh}), we have performed a particle-hole transformation, and $c^\dagger_{i \sigma}$ is the creation operator of a hole at site $i$ with spin $\sigma$. The phase of the next-nearest-neighbor hopping is $\pm\nu_{ij}\theta$ ($+$ for spin up holes), where $\nu_{ij}$ depends on the hopping directions [Fig.~\ref{fig:fig1}(a)]. If the top four moir\'e bands are well separated from the remaining bands, $\theta = \pi/3$ can be deduced from momentum mismatch via Peierls substitution \cite{wu2019topological}. The phase diagram of Kane-Mele-Hubbard model as a function of $U$ and $\Delta_D$ is first obtained in Ref.~\cite{jiang2018antiferromagnetic} for $\theta=\pi / 2$. In Fig.~\ref{fig:KMH}(a), we obtain a similar phase diagram for $\theta=\pi / 3$. We find the phase diagrams for the continuum model and the Kane-Mele-Hubbard model are qualitatively similar. In the following, analysis based on Kane-Mele-Hubbard model will be presented to understand the emergence of correlated phases in tMoTe$_2$.

\textit{Spin model at $\Delta_D=0$.}--- At $\nu = -2$, without the electric field, each site is occupied by one electron. In the strong coupling regime, the Kane-Mele-Hubbard model can be projected into a spin model \cite{hohenadler2012quantum,rachel2010topological,zare2021spin}
\begin{equation}\label{eq:spin}
\begin{aligned}
    H_{\rm spin} = & ~J_1 \sum_{\langle i, j\rangle} \bm{S}_i \cdot \bm{S}_j + 
    \sum_{\langle\langle i, j \rangle\rangle} J_2^z S_i^z S_j^z \\
    & + J_2^{xy} (S_i^x S_j^x + S_i^y S_j^y) + D (\bm{S}_i\times \bm{S}_j) \cdot \hat{\bm{z}}, 
\end{aligned}
\end{equation}
with $J_1=4t_1^2/U$ , $J_2^z = 4 t_2^2/U$, $J_2^{xy} = 4t_2^2 \cos(2\theta)/U$ and $D = 4t_2^2 \sin(2\theta)/U$. The direction for $\langle\langle i, j\rangle\rangle$ follows the arrows in Fig.~\ref{fig:fig1}(a).  Therefore, $J_1$ is always positive, leading to antiferromagnetic interactions between sublattices. The sign of $J_2^{xy}$ depends on the $\theta$. For $\theta = \pi / 3$, $J_2^{xy} < 0$, leading to the AFMxy phase in the Hartree-Fock phase diagram. For other values of $\theta$, translational symmetry breaking phase and spin liquid phase may appear \cite{zare2021spin}.

\textit{The emergence of the AFMz phase.}---
To explain the existence of the AFMz phase, we find it useful to investigate the behavior of an infinitesimal antiferromagnetic perturbation to the phase boundary between the QSH and the BI in the mean field theory. On the QSH-BI phase boundary, the quasiparticle band gap closes at the $\bm{\kappa_+}$ and $\bm{\kappa_-}$ points, forming Dirac points [Fig.~\ref{fig:KMH}(b)]. An antiferromagnetic perturbation in the $z$ direction (AFMz perturbation) will open up gaps at both $\bm{\kappa_+}$ and $\bm{\kappa_-}$ points, providing an opportunity to lower the energy around the Dirac points [Fig.~\ref{fig:KMH}(b)]. We identify the gap opening as the driving force in stabilizing the AFMz phase at the QSH-BI phase boundary.

As only onsite Coulomb interactions are considered in the Kane-Mele-Hubbard model, the local magnetic moments completely determine the Hartree-Fock quasiparticle Hamiltonian $H_{\mathrm{HF}}$. In AFMz, QSH and BI phases, our simulations find no in-plane magnetic components, thus the mean-field decoupled Hartree-Fock Hamiltonian can be written as $H_{\mathrm{HF}} = H_{\mathrm{single}} + U \sum_{i,\sigma} n_{i \bar{\sigma}} c^\dagger_{i\sigma} c_{i\sigma}$, where $n_{i\bar{\sigma}}$ is the occupation number at site $i$ with spin $-\sigma$. Due to translational symmetry of the AFMz, QSH and BI states, $n_{i\sigma}$ only depends on the sublattice and spin, and takes value $n_{A\sigma}$ or $n_{B\sigma}$. An AFMz perturbation $\delta n_{A\uparrow} = \delta n_{B\downarrow} = -\delta n_{A\downarrow} = -\delta n_{B\uparrow} = \delta n$ will induce a perturbative Hartree-Fock Hamiltonian $\delta H_{\mathrm{HF}}$. Accordingly, the quasiparticle energies $\epsilon_\alpha(\bm{k})$ will receive a perturbative correction $\delta \epsilon_\alpha(\bm{k})$ ($\alpha$ is the band index). For a single band that are well separated from other bands by an energy gap, $ \delta \epsilon_\alpha(\bm{k}) \approx - \delta \epsilon_\alpha(\bm{-k})$, because $\delta \hat{H}_{\mathrm{HF}}$ is odd under time reversal symmetry~\footnote{The argument can be easily generalized to summations of band energies of nearly degenerate bands}. For example, the energy of the lowest band in Fig.~\ref{fig:KMH}(b) shifts in opposite directions at $\bm{\kappa_-}$ and $\bm{\kappa_+}$ under an AFMz perturbation (red vs. black lines). Therefore, the change of the total quasiparticle energy $\delta E_\mathrm{QP} = \sum_{\bm{k},\alpha \in \mathrm{occupied}} \delta\epsilon_\alpha(\bm{k})$ will be mainly contributed by the $\bm{k}$ points where the quasiparticle energies might shift above or below the Fermi energy. In particular, at the QSH-BI phase boundary, $\delta E_\mathrm{QP}$ is dominated by the gap opening at the Dirac points [Fig.~\ref{fig:KMH}(b)]. The total quasiparticle energy $E_\mathrm{QP}$ has double counted the interaction energy. The correct energy of a Hartree-Fock state is $E = E_\mathrm{QP} - E_\mathrm{int}$, where the interaction energy $E_\mathrm{int} = U n_{i\uparrow} n_{i\downarrow}$ for states without in-plane magnetic moments. For an infinitesimal AFMz perturbation to the QSH-BI phase boundary, $\delta E_\mathrm{QP} \propto -U \delta n$ (the opened gap is proportional to $U \delta n$), while the change of interaction energy $\delta E_\mathrm{int} \propto  -U \delta n^2$ ($n_{i\uparrow} = n_{i\downarrow}$ for nonmagnetic states). Therefore, an infinitesimal AFMz perturbation will always lower energy at the QSH-BI phase boundary. However, the AFMz perturbation itself has to be consistently enforced. Especially, the interaction has to be sufficiently large, such that local magnetic moments can be stabilized. This is evidenced in Fig.~\ref{fig:KMH}(a), where a critical $U$ is needed for the existence of the AFMz phase.

Finally, at the QSH-BI phase boundary, an antiferromagnetic perturbation in the $x-y$ plane (AFMxy perturbation) will open up a much smaller gap at the Dirac points compared to an AFMz perturbation in most cases. This is because an AFMxy perturbation has no first order perturbative effects to the Dirac points due to the spin-split quasiparticle band structure [Fig.~\ref{fig:KMH}(b)] unless $\theta = 0$ or $\pi$, for which there is no spin orbit interaction and the bands are spin degenerate.

\begin{figure}
\centering
\includegraphics[width=0.617\columnwidth]{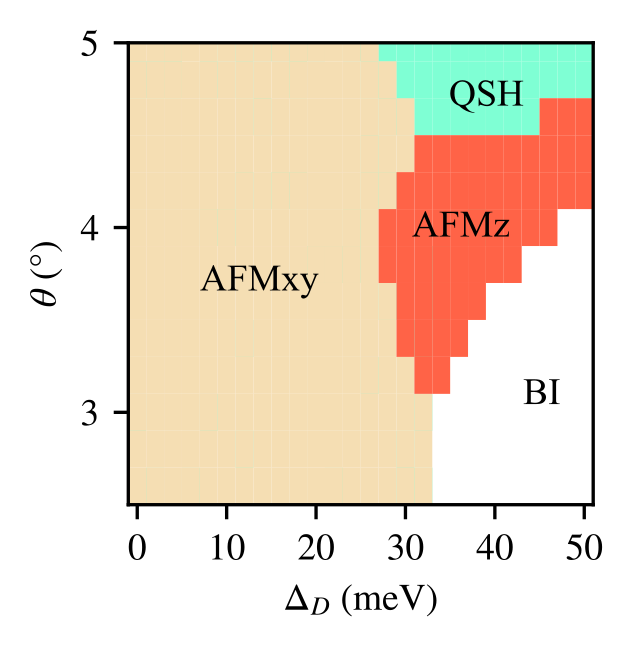}
\caption{The phase diagram of the continuum model at $\epsilon = 20$. \label{fig:angle}}
\end{figure}

\textit{Twist angle dependence.}---
The twist angle changes the period of the moir\'e potential and serves as another tuning knob in moir\'e systems. In Fig.~\ref{fig:angle}, we present the phase diagram as a function of the twist angle and $\Delta_D$. Deviations from Kane-Mele-Hubbard model appears for some twist angles. Especially, for small twist angles, there is a phase boundary between AFMxy and BI phase, which does not appear in the phase diagram of the Kane-Mele-Hubbard model. One cause is the reduction of $t_2$ with respect to $t_1$. In the Supplemental Material~\cite{supp}, we show that for small $t_2$, the area of the AFMz phase shrinks significantly in the phase diagram of Kane-Mele-Hubbard model. Therefore, AFMz phase might completely disappear if further ingredients are added to the Kane-Mele-Hubbard model, explaining the absence of the AFMz state for $\epsilon = 20$ and $\theta < 3^\circ$ for the continuum model. For large twist angles, both AFMz and AFMxy are gradually suppressed due to large band dispersion. We note that our calculations are most accurate around $~3.89^\circ$ where the continuum model parameters are extracted from the density functional theory calculations.

In summary, we have discovered an antiferromagnetic Chern state in twisted bilayer MoTe$_2$ at hole doping $\nu = -2$ under an out-of-plane electric field. Experimentally, due to vanishing total magnetic moment, a direct or indirect measurement of Hall conductivity is required to confirm the existence of state. The easiest route is probably inferring the Hall conductivity by Streda formula \cite{cai2023signatures}, which requires the use of an external magnetic field that does not alter the system's magnetic state. For a spin model, where the local magnetic moments have fixed magnitudes, the stability of an antiferromagnetic phase under external magnetic field is usually determined by magnetic anisotropy. In the current case, however, the magnitudes of local magnetic moments are not fixed, and the antiferromagnetic Chern state do not admit a straightforward spin model description. Further investigations are needed to determine the stability of the antiferromagnetic Chern state with respect to an external magnetic field.

\begin{acknowledgments}
We thank Xiaodong Xu and Yuchi He for stimulating discussions.  This work is mainly supported by the Center on Programmable Quantum Materials, an Energy Frontier Research Center funded by DOE BES under award DE-SC0019443.  The real-space Hartree-Fock calculation is supported by the Department of Energy BES QIS program under award DE-SC0022277.  This work was facilitated through the use of advanced computational, storage, and networking infrastructure provided by the Hyak supercomputer system and funded by the University of Washington Molecular Engineering Materials Center at the University of Washington (NSF MRSEC DMR-1719797).
\end{acknowledgments}

\begin{acknowledgments}

\end{acknowledgments}

\bibliography{main}

\begin{thebibliography}{42}%
\makeatletter
\providecommand \@ifxundefined [1]{%
 \@ifx{#1\undefined}
}%
\providecommand \@ifnum [1]{%
 \ifnum #1\expandafter \@firstoftwo
 \else \expandafter \@secondoftwo
 \fi
}%
\providecommand \@ifx [1]{%
 \ifx #1\expandafter \@firstoftwo
 \else \expandafter \@secondoftwo
 \fi
}%
\providecommand \natexlab [1]{#1}%
\providecommand \enquote  [1]{``#1''}%
\providecommand \bibnamefont  [1]{#1}%
\providecommand \bibfnamefont [1]{#1}%
\providecommand \citenamefont [1]{#1}%
\providecommand \href@noop [0]{\@secondoftwo}%
\providecommand \href [0]{\begingroup \@sanitize@url \@href}%
\providecommand \@href[1]{\@@startlink{#1}\@@href}%
\providecommand \@@href[1]{\endgroup#1\@@endlink}%
\providecommand \@sanitize@url [0]{\catcode `\\12\catcode `\$12\catcode
  `\&12\catcode `\#12\catcode `\^12\catcode `\_12\catcode `\%12\relax}%
\providecommand \@@startlink[1]{}%
\providecommand \@@endlink[0]{}%
\providecommand \url  [0]{\begingroup\@sanitize@url \@url }%
\providecommand \@url [1]{\endgroup\@href {#1}{\urlprefix }}%
\providecommand \urlprefix  [0]{URL }%
\providecommand \Eprint [0]{\href }%
\providecommand \doibase [0]{https://doi.org/}%
\providecommand \selectlanguage [0]{\@gobble}%
\providecommand \bibinfo  [0]{\@secondoftwo}%
\providecommand \bibfield  [0]{\@secondoftwo}%
\providecommand \translation [1]{[#1]}%
\providecommand \BibitemOpen [0]{}%
\providecommand \bibitemStop [0]{}%
\providecommand \bibitemNoStop [0]{.\EOS\space}%
\providecommand \EOS [0]{\spacefactor3000\relax}%
\providecommand \BibitemShut  [1]{\csname bibitem#1\endcsname}%
\let\auto@bib@innerbib\@empty
%</preamble>
\bibitem [{\citenamefont {Cao}\ \emph {et~al.}(2018{\natexlab{a}})\citenamefont
  {Cao}, \citenamefont {Fatemi}, \citenamefont {Fang}, \citenamefont
  {Watanabe}, \citenamefont {Taniguchi}, \citenamefont {Kaxiras},\ and\
  \citenamefont {Jarillo-Herrero}}]{cao2018unconventional}%
  \BibitemOpen
  \bibfield  {author} {\bibinfo {author} {\bibfnamefont {Y.}~\bibnamefont
  {Cao}}, \bibinfo {author} {\bibfnamefont {V.}~\bibnamefont {Fatemi}},
  \bibinfo {author} {\bibfnamefont {S.}~\bibnamefont {Fang}}, \bibinfo {author}
  {\bibfnamefont {K.}~\bibnamefont {Watanabe}}, \bibinfo {author}
  {\bibfnamefont {T.}~\bibnamefont {Taniguchi}}, \bibinfo {author}
  {\bibfnamefont {E.}~\bibnamefont {Kaxiras}},\ and\ \bibinfo {author}
  {\bibfnamefont {P.}~\bibnamefont {Jarillo-Herrero}},\ }\bibfield  {title}
  {\bibinfo {title} {Unconventional superconductivity in magic-angle graphene
  superlattices},\ }\href@noop {} {\bibfield  {journal} {\bibinfo  {journal}
  {Nature}\ }\textbf {\bibinfo {volume} {556}},\ \bibinfo {pages} {43}
  (\bibinfo {year} {2018}{\natexlab{a}})}\BibitemShut {NoStop}%
\bibitem [{\citenamefont {Cao}\ \emph {et~al.}(2018{\natexlab{b}})\citenamefont
  {Cao}, \citenamefont {Fatemi}, \citenamefont {Demir}, \citenamefont {Fang},
  \citenamefont {Tomarken}, \citenamefont {Luo}, \citenamefont
  {Sanchez-Yamagishi}, \citenamefont {Watanabe}, \citenamefont {Taniguchi},
  \citenamefont {Kaxiras} \emph {et~al.}}]{cao2018correlated}%
  \BibitemOpen
  \bibfield  {author} {\bibinfo {author} {\bibfnamefont {Y.}~\bibnamefont
  {Cao}}, \bibinfo {author} {\bibfnamefont {V.}~\bibnamefont {Fatemi}},
  \bibinfo {author} {\bibfnamefont {A.}~\bibnamefont {Demir}}, \bibinfo
  {author} {\bibfnamefont {S.}~\bibnamefont {Fang}}, \bibinfo {author}
  {\bibfnamefont {S.~L.}\ \bibnamefont {Tomarken}}, \bibinfo {author}
  {\bibfnamefont {J.~Y.}\ \bibnamefont {Luo}}, \bibinfo {author} {\bibfnamefont
  {J.~D.}\ \bibnamefont {Sanchez-Yamagishi}}, \bibinfo {author} {\bibfnamefont
  {K.}~\bibnamefont {Watanabe}}, \bibinfo {author} {\bibfnamefont
  {T.}~\bibnamefont {Taniguchi}}, \bibinfo {author} {\bibfnamefont
  {E.}~\bibnamefont {Kaxiras}}, \emph {et~al.},\ }\bibfield  {title} {\bibinfo
  {title} {Correlated insulator behaviour at half-filling in magic-angle
  graphene superlattices},\ }\href@noop {} {\bibfield  {journal} {\bibinfo
  {journal} {Nature}\ }\textbf {\bibinfo {volume} {556}},\ \bibinfo {pages}
  {80} (\bibinfo {year} {2018}{\natexlab{b}})}\BibitemShut {NoStop}%
\bibitem [{\citenamefont {Wu}\ \emph {et~al.}(2018)\citenamefont {Wu},
  \citenamefont {Lovorn}, \citenamefont {Tutuc},\ and\ \citenamefont
  {MacDonald}}]{wu2018hubbard}%
  \BibitemOpen
  \bibfield  {author} {\bibinfo {author} {\bibfnamefont {F.}~\bibnamefont
  {Wu}}, \bibinfo {author} {\bibfnamefont {T.}~\bibnamefont {Lovorn}}, \bibinfo
  {author} {\bibfnamefont {E.}~\bibnamefont {Tutuc}},\ and\ \bibinfo {author}
  {\bibfnamefont {A.~H.}\ \bibnamefont {MacDonald}},\ }\bibfield  {title}
  {\bibinfo {title} {Hubbard model physics in transition metal dichalcogenide
  moir{\'e} bands},\ }\href@noop {} {\bibfield  {journal} {\bibinfo  {journal}
  {Phys. Rev. Lett.}\ }\textbf {\bibinfo {volume} {121}},\ \bibinfo {pages}
  {026402} (\bibinfo {year} {2018})}\BibitemShut {NoStop}%
\bibitem [{\citenamefont {Wu}\ \emph {et~al.}(2019)\citenamefont {Wu},
  \citenamefont {Lovorn}, \citenamefont {Tutuc}, \citenamefont {Martin},\ and\
  \citenamefont {MacDonald}}]{wu2019topological}%
  \BibitemOpen
  \bibfield  {author} {\bibinfo {author} {\bibfnamefont {F.}~\bibnamefont
  {Wu}}, \bibinfo {author} {\bibfnamefont {T.}~\bibnamefont {Lovorn}}, \bibinfo
  {author} {\bibfnamefont {E.}~\bibnamefont {Tutuc}}, \bibinfo {author}
  {\bibfnamefont {I.}~\bibnamefont {Martin}},\ and\ \bibinfo {author}
  {\bibfnamefont {A.}~\bibnamefont {MacDonald}},\ }\bibfield  {title} {\bibinfo
  {title} {Topological insulators in twisted transition metal dichalcogenide
  homobilayers},\ }\href@noop {} {\bibfield  {journal} {\bibinfo  {journal}
  {Phys. Rev. Lett.}\ }\textbf {\bibinfo {volume} {122}},\ \bibinfo {pages}
  {086402} (\bibinfo {year} {2019})}\BibitemShut {NoStop}%
\bibitem [{\citenamefont {Regan}\ \emph {et~al.}(2020)\citenamefont {Regan},
  \citenamefont {Wang}, \citenamefont {Jin}, \citenamefont {Bakti~Utama},
  \citenamefont {Gao}, \citenamefont {Wei}, \citenamefont {Zhao}, \citenamefont
  {Zhao}, \citenamefont {Zhang}, \citenamefont {Yumigeta} \emph
  {et~al.}}]{regan2020mott}%
  \BibitemOpen
  \bibfield  {author} {\bibinfo {author} {\bibfnamefont {E.~C.}\ \bibnamefont
  {Regan}}, \bibinfo {author} {\bibfnamefont {D.}~\bibnamefont {Wang}},
  \bibinfo {author} {\bibfnamefont {C.}~\bibnamefont {Jin}}, \bibinfo {author}
  {\bibfnamefont {M.~I.}\ \bibnamefont {Bakti~Utama}}, \bibinfo {author}
  {\bibfnamefont {B.}~\bibnamefont {Gao}}, \bibinfo {author} {\bibfnamefont
  {X.}~\bibnamefont {Wei}}, \bibinfo {author} {\bibfnamefont {S.}~\bibnamefont
  {Zhao}}, \bibinfo {author} {\bibfnamefont {W.}~\bibnamefont {Zhao}}, \bibinfo
  {author} {\bibfnamefont {Z.}~\bibnamefont {Zhang}}, \bibinfo {author}
  {\bibfnamefont {K.}~\bibnamefont {Yumigeta}}, \emph {et~al.},\ }\bibfield
  {title} {\bibinfo {title} {Mott and generalized wigner crystal states in
  {WSe2/WS2} moir{\'e} superlattices},\ }\href@noop {} {\bibfield  {journal}
  {\bibinfo  {journal} {Nature}\ }\textbf {\bibinfo {volume} {579}},\ \bibinfo
  {pages} {359} (\bibinfo {year} {2020})}\BibitemShut {NoStop}%
\bibitem [{\citenamefont {Tang}\ \emph {et~al.}(2020)\citenamefont {Tang},
  \citenamefont {Li}, \citenamefont {Li}, \citenamefont {Xu}, \citenamefont
  {Liu}, \citenamefont {Barmak}, \citenamefont {Watanabe}, \citenamefont
  {Taniguchi}, \citenamefont {MacDonald}, \citenamefont {Shan} \emph
  {et~al.}}]{tang2020simulation}%
  \BibitemOpen
  \bibfield  {author} {\bibinfo {author} {\bibfnamefont {Y.}~\bibnamefont
  {Tang}}, \bibinfo {author} {\bibfnamefont {L.}~\bibnamefont {Li}}, \bibinfo
  {author} {\bibfnamefont {T.}~\bibnamefont {Li}}, \bibinfo {author}
  {\bibfnamefont {Y.}~\bibnamefont {Xu}}, \bibinfo {author} {\bibfnamefont
  {S.}~\bibnamefont {Liu}}, \bibinfo {author} {\bibfnamefont {K.}~\bibnamefont
  {Barmak}}, \bibinfo {author} {\bibfnamefont {K.}~\bibnamefont {Watanabe}},
  \bibinfo {author} {\bibfnamefont {T.}~\bibnamefont {Taniguchi}}, \bibinfo
  {author} {\bibfnamefont {A.~H.}\ \bibnamefont {MacDonald}}, \bibinfo {author}
  {\bibfnamefont {J.}~\bibnamefont {Shan}}, \emph {et~al.},\ }\bibfield
  {title} {\bibinfo {title} {Simulation of hubbard model physics in {WSe2/WS2}
  moir{\'e} superlattices},\ }\href@noop {} {\bibfield  {journal} {\bibinfo
  {journal} {Nature}\ }\textbf {\bibinfo {volume} {579}},\ \bibinfo {pages}
  {353} (\bibinfo {year} {2020})}\BibitemShut {NoStop}%
\bibitem [{\citenamefont {Zhao}\ \emph {et~al.}(2023)\citenamefont {Zhao},
  \citenamefont {Shen}, \citenamefont {Tao}, \citenamefont {Han}, \citenamefont
  {Kang}, \citenamefont {Watanabe}, \citenamefont {Taniguchi}, \citenamefont
  {Mak},\ and\ \citenamefont {Shan}}]{zhao2023gate}%
  \BibitemOpen
  \bibfield  {author} {\bibinfo {author} {\bibfnamefont {W.}~\bibnamefont
  {Zhao}}, \bibinfo {author} {\bibfnamefont {B.}~\bibnamefont {Shen}}, \bibinfo
  {author} {\bibfnamefont {Z.}~\bibnamefont {Tao}}, \bibinfo {author}
  {\bibfnamefont {Z.}~\bibnamefont {Han}}, \bibinfo {author} {\bibfnamefont
  {K.}~\bibnamefont {Kang}}, \bibinfo {author} {\bibfnamefont {K.}~\bibnamefont
  {Watanabe}}, \bibinfo {author} {\bibfnamefont {T.}~\bibnamefont {Taniguchi}},
  \bibinfo {author} {\bibfnamefont {K.~F.}\ \bibnamefont {Mak}},\ and\ \bibinfo
  {author} {\bibfnamefont {J.}~\bibnamefont {Shan}},\ }\bibfield  {title}
  {\bibinfo {title} {Gate-tunable heavy fermions in a moir{\'e} kondo
  lattice},\ }\href@noop {} {\bibfield  {journal} {\bibinfo  {journal}
  {Nature}\ }\textbf {\bibinfo {volume} {616}},\ \bibinfo {pages} {61}
  (\bibinfo {year} {2023})}\BibitemShut {NoStop}%
\bibitem [{\citenamefont {Li}\ \emph {et~al.}(2021{\natexlab{a}})\citenamefont
  {Li}, \citenamefont {Jiang}, \citenamefont {Li}, \citenamefont {Zhang},
  \citenamefont {Kang}, \citenamefont {Zhu}, \citenamefont {Watanabe},
  \citenamefont {Taniguchi}, \citenamefont {Chowdhury}, \citenamefont {Fu}
  \emph {et~al.}}]{li2021continuous}%
  \BibitemOpen
  \bibfield  {author} {\bibinfo {author} {\bibfnamefont {T.}~\bibnamefont
  {Li}}, \bibinfo {author} {\bibfnamefont {S.}~\bibnamefont {Jiang}}, \bibinfo
  {author} {\bibfnamefont {L.}~\bibnamefont {Li}}, \bibinfo {author}
  {\bibfnamefont {Y.}~\bibnamefont {Zhang}}, \bibinfo {author} {\bibfnamefont
  {K.}~\bibnamefont {Kang}}, \bibinfo {author} {\bibfnamefont {J.}~\bibnamefont
  {Zhu}}, \bibinfo {author} {\bibfnamefont {K.}~\bibnamefont {Watanabe}},
  \bibinfo {author} {\bibfnamefont {T.}~\bibnamefont {Taniguchi}}, \bibinfo
  {author} {\bibfnamefont {D.}~\bibnamefont {Chowdhury}}, \bibinfo {author}
  {\bibfnamefont {L.}~\bibnamefont {Fu}}, \emph {et~al.},\ }\bibfield  {title}
  {\bibinfo {title} {Continuous {Mott} transition in semiconductor moir{\'e}
  superlattices},\ }\href@noop {} {\bibfield  {journal} {\bibinfo  {journal}
  {Nature}\ }\textbf {\bibinfo {volume} {597}},\ \bibinfo {pages} {350}
  (\bibinfo {year} {2021}{\natexlab{a}})}\BibitemShut {NoStop}%
\bibitem [{\citenamefont {Xu}\ \emph {et~al.}(2020)\citenamefont {Xu},
  \citenamefont {Liu}, \citenamefont {Rhodes}, \citenamefont {Watanabe},
  \citenamefont {Taniguchi}, \citenamefont {Hone}, \citenamefont {Elser},
  \citenamefont {Mak},\ and\ \citenamefont {Shan}}]{xu2020correlated}%
  \BibitemOpen
  \bibfield  {author} {\bibinfo {author} {\bibfnamefont {Y.}~\bibnamefont
  {Xu}}, \bibinfo {author} {\bibfnamefont {S.}~\bibnamefont {Liu}}, \bibinfo
  {author} {\bibfnamefont {D.~A.}\ \bibnamefont {Rhodes}}, \bibinfo {author}
  {\bibfnamefont {K.}~\bibnamefont {Watanabe}}, \bibinfo {author}
  {\bibfnamefont {T.}~\bibnamefont {Taniguchi}}, \bibinfo {author}
  {\bibfnamefont {J.}~\bibnamefont {Hone}}, \bibinfo {author} {\bibfnamefont
  {V.}~\bibnamefont {Elser}}, \bibinfo {author} {\bibfnamefont {K.~F.}\
  \bibnamefont {Mak}},\ and\ \bibinfo {author} {\bibfnamefont {J.}~\bibnamefont
  {Shan}},\ }\bibfield  {title} {\bibinfo {title} {Correlated insulating states
  at fractional fillings of moir{\'e} superlattices},\ }\href@noop {}
  {\bibfield  {journal} {\bibinfo  {journal} {Nature}\ }\textbf {\bibinfo
  {volume} {587}},\ \bibinfo {pages} {214} (\bibinfo {year}
  {2020})}\BibitemShut {NoStop}%
\bibitem [{\citenamefont {Huang}\ \emph {et~al.}(2021)\citenamefont {Huang},
  \citenamefont {Wang}, \citenamefont {Miao}, \citenamefont {Wang},
  \citenamefont {Li}, \citenamefont {Lian}, \citenamefont {Taniguchi},
  \citenamefont {Watanabe}, \citenamefont {Okamoto}, \citenamefont {Xiao} \emph
  {et~al.}}]{huang2021correlated}%
  \BibitemOpen
  \bibfield  {author} {\bibinfo {author} {\bibfnamefont {X.}~\bibnamefont
  {Huang}}, \bibinfo {author} {\bibfnamefont {T.}~\bibnamefont {Wang}},
  \bibinfo {author} {\bibfnamefont {S.}~\bibnamefont {Miao}}, \bibinfo {author}
  {\bibfnamefont {C.}~\bibnamefont {Wang}}, \bibinfo {author} {\bibfnamefont
  {Z.}~\bibnamefont {Li}}, \bibinfo {author} {\bibfnamefont {Z.}~\bibnamefont
  {Lian}}, \bibinfo {author} {\bibfnamefont {T.}~\bibnamefont {Taniguchi}},
  \bibinfo {author} {\bibfnamefont {K.}~\bibnamefont {Watanabe}}, \bibinfo
  {author} {\bibfnamefont {S.}~\bibnamefont {Okamoto}}, \bibinfo {author}
  {\bibfnamefont {D.}~\bibnamefont {Xiao}}, \emph {et~al.},\ }\bibfield
  {title} {\bibinfo {title} {Correlated insulating states at fractional
  fillings of the {WS2/WSe2} moir{\'e} lattice},\ }\href@noop {} {\bibfield
  {journal} {\bibinfo  {journal} {Nat. Phys.}\ }\textbf {\bibinfo {volume}
  {17}},\ \bibinfo {pages} {715} (\bibinfo {year} {2021})}\BibitemShut
  {NoStop}%
\bibitem [{\citenamefont {Li}\ \emph {et~al.}(2021{\natexlab{b}})\citenamefont
  {Li}, \citenamefont {Li}, \citenamefont {Regan}, \citenamefont {Wang},
  \citenamefont {Zhao}, \citenamefont {Kahn}, \citenamefont {Yumigeta},
  \citenamefont {Blei}, \citenamefont {Taniguchi}, \citenamefont {Watanabe}
  \emph {et~al.}}]{li2021imaging}%
  \BibitemOpen
  \bibfield  {author} {\bibinfo {author} {\bibfnamefont {H.}~\bibnamefont
  {Li}}, \bibinfo {author} {\bibfnamefont {S.}~\bibnamefont {Li}}, \bibinfo
  {author} {\bibfnamefont {E.~C.}\ \bibnamefont {Regan}}, \bibinfo {author}
  {\bibfnamefont {D.}~\bibnamefont {Wang}}, \bibinfo {author} {\bibfnamefont
  {W.}~\bibnamefont {Zhao}}, \bibinfo {author} {\bibfnamefont {S.}~\bibnamefont
  {Kahn}}, \bibinfo {author} {\bibfnamefont {K.}~\bibnamefont {Yumigeta}},
  \bibinfo {author} {\bibfnamefont {M.}~\bibnamefont {Blei}}, \bibinfo {author}
  {\bibfnamefont {T.}~\bibnamefont {Taniguchi}}, \bibinfo {author}
  {\bibfnamefont {K.}~\bibnamefont {Watanabe}}, \emph {et~al.},\ }\bibfield
  {title} {\bibinfo {title} {Imaging two-dimensional generalized {Wigner}
  crystals},\ }\href@noop {} {\bibfield  {journal} {\bibinfo  {journal}
  {Nature}\ }\textbf {\bibinfo {volume} {597}},\ \bibinfo {pages} {650}
  (\bibinfo {year} {2021}{\natexlab{b}})}\BibitemShut {NoStop}%
\bibitem [{\citenamefont {Li}\ \emph {et~al.}(2021{\natexlab{c}})\citenamefont
  {Li}, \citenamefont {Jiang}, \citenamefont {Shen}, \citenamefont {Zhang},
  \citenamefont {Li}, \citenamefont {Tao}, \citenamefont {Devakul},
  \citenamefont {Watanabe}, \citenamefont {Taniguchi}, \citenamefont {Fu} \emph
  {et~al.}}]{li2021quantum}%
  \BibitemOpen
  \bibfield  {author} {\bibinfo {author} {\bibfnamefont {T.}~\bibnamefont
  {Li}}, \bibinfo {author} {\bibfnamefont {S.}~\bibnamefont {Jiang}}, \bibinfo
  {author} {\bibfnamefont {B.}~\bibnamefont {Shen}}, \bibinfo {author}
  {\bibfnamefont {Y.}~\bibnamefont {Zhang}}, \bibinfo {author} {\bibfnamefont
  {L.}~\bibnamefont {Li}}, \bibinfo {author} {\bibfnamefont {Z.}~\bibnamefont
  {Tao}}, \bibinfo {author} {\bibfnamefont {T.}~\bibnamefont {Devakul}},
  \bibinfo {author} {\bibfnamefont {K.}~\bibnamefont {Watanabe}}, \bibinfo
  {author} {\bibfnamefont {T.}~\bibnamefont {Taniguchi}}, \bibinfo {author}
  {\bibfnamefont {L.}~\bibnamefont {Fu}}, \emph {et~al.},\ }\bibfield  {title}
  {\bibinfo {title} {Quantum anomalous hall effect from intertwined moir{\'e}
  bands},\ }\href@noop {} {\bibfield  {journal} {\bibinfo  {journal} {Nature}\
  }\textbf {\bibinfo {volume} {600}},\ \bibinfo {pages} {641} (\bibinfo {year}
  {2021}{\natexlab{c}})}\BibitemShut {NoStop}%
\bibitem [{\citenamefont {Tao}\ \emph {et~al.}(2022)\citenamefont {Tao},
  \citenamefont {Shen}, \citenamefont {Jiang}, \citenamefont {Li},
  \citenamefont {Li}, \citenamefont {Ma}, \citenamefont {Zhao}, \citenamefont
  {Hu}, \citenamefont {Pistunova}, \citenamefont {Watanabe} \emph
  {et~al.}}]{tao2022valley}%
  \BibitemOpen
  \bibfield  {author} {\bibinfo {author} {\bibfnamefont {Z.}~\bibnamefont
  {Tao}}, \bibinfo {author} {\bibfnamefont {B.}~\bibnamefont {Shen}}, \bibinfo
  {author} {\bibfnamefont {S.}~\bibnamefont {Jiang}}, \bibinfo {author}
  {\bibfnamefont {T.}~\bibnamefont {Li}}, \bibinfo {author} {\bibfnamefont
  {L.}~\bibnamefont {Li}}, \bibinfo {author} {\bibfnamefont {L.}~\bibnamefont
  {Ma}}, \bibinfo {author} {\bibfnamefont {W.}~\bibnamefont {Zhao}}, \bibinfo
  {author} {\bibfnamefont {J.}~\bibnamefont {Hu}}, \bibinfo {author}
  {\bibfnamefont {K.}~\bibnamefont {Pistunova}}, \bibinfo {author}
  {\bibfnamefont {K.}~\bibnamefont {Watanabe}}, \emph {et~al.},\ }\bibfield
  {title} {\bibinfo {title} {Valley-coherent quantum anomalous hall state in
  ab-stacked mote2/wse2 bilayers},\ }\href@noop {} {\bibfield  {journal}
  {\bibinfo  {journal} {arXiv:2208.07452}\ } (\bibinfo {year}
  {2022})}\BibitemShut {NoStop}%
\bibitem [{\citenamefont {Zhao}\ \emph {et~al.}(2022)\citenamefont {Zhao},
  \citenamefont {Kang}, \citenamefont {Li}, \citenamefont {Tschirhart},
  \citenamefont {Redekop}, \citenamefont {Watanabe}, \citenamefont {Taniguchi},
  \citenamefont {Young}, \citenamefont {Shan},\ and\ \citenamefont
  {Mak}}]{zhao2022realization}%
  \BibitemOpen
  \bibfield  {author} {\bibinfo {author} {\bibfnamefont {W.}~\bibnamefont
  {Zhao}}, \bibinfo {author} {\bibfnamefont {K.}~\bibnamefont {Kang}}, \bibinfo
  {author} {\bibfnamefont {L.}~\bibnamefont {Li}}, \bibinfo {author}
  {\bibfnamefont {C.}~\bibnamefont {Tschirhart}}, \bibinfo {author}
  {\bibfnamefont {E.}~\bibnamefont {Redekop}}, \bibinfo {author} {\bibfnamefont
  {K.}~\bibnamefont {Watanabe}}, \bibinfo {author} {\bibfnamefont
  {T.}~\bibnamefont {Taniguchi}}, \bibinfo {author} {\bibfnamefont
  {A.}~\bibnamefont {Young}}, \bibinfo {author} {\bibfnamefont
  {J.}~\bibnamefont {Shan}},\ and\ \bibinfo {author} {\bibfnamefont {K.~F.}\
  \bibnamefont {Mak}},\ }\bibfield  {title} {\bibinfo {title} {Realization of
  the {Haldane Chern} insulator in a moir{\'e} lattice},\ }\href@noop {}
  {\bibfield  {journal} {\bibinfo  {journal} {arXiv:2207.02312}\ } (\bibinfo
  {year} {2022})}\BibitemShut {NoStop}%
\bibitem [{\citenamefont {Zhang}\ \emph {et~al.}(2021)\citenamefont {Zhang},
  \citenamefont {Devakul},\ and\ \citenamefont {Fu}}]{zhang2021spin}%
  \BibitemOpen
  \bibfield  {author} {\bibinfo {author} {\bibfnamefont {Y.}~\bibnamefont
  {Zhang}}, \bibinfo {author} {\bibfnamefont {T.}~\bibnamefont {Devakul}},\
  and\ \bibinfo {author} {\bibfnamefont {L.}~\bibnamefont {Fu}},\ }\bibfield
  {title} {\bibinfo {title} {Spin-textured chern bands in ab-stacked transition
  metal dichalcogenide bilayers},\ }\href@noop {} {\bibfield  {journal}
  {\bibinfo  {journal} {Proc. Natl. Acad. Sci. U.S.A.}\ }\textbf {\bibinfo
  {volume} {118}},\ \bibinfo {pages} {e2112673118} (\bibinfo {year}
  {2021})}\BibitemShut {NoStop}%
\bibitem [{\citenamefont {Devakul}\ \emph {et~al.}(2021)\citenamefont
  {Devakul}, \citenamefont {Cr{\'e}pel}, \citenamefont {Zhang},\ and\
  \citenamefont {Fu}}]{devakul2021magic}%
  \BibitemOpen
  \bibfield  {author} {\bibinfo {author} {\bibfnamefont {T.}~\bibnamefont
  {Devakul}}, \bibinfo {author} {\bibfnamefont {V.}~\bibnamefont {Cr{\'e}pel}},
  \bibinfo {author} {\bibfnamefont {Y.}~\bibnamefont {Zhang}},\ and\ \bibinfo
  {author} {\bibfnamefont {L.}~\bibnamefont {Fu}},\ }\bibfield  {title}
  {\bibinfo {title} {Magic in twisted transition metal dichalcogenide
  bilayers},\ }\href@noop {} {\bibfield  {journal} {\bibinfo  {journal} {Nat.
  Commun.}\ }\textbf {\bibinfo {volume} {12}},\ \bibinfo {pages} {6730}
  (\bibinfo {year} {2021})}\BibitemShut {NoStop}%
\bibitem [{\citenamefont {Pan}\ \emph {et~al.}(2022)\citenamefont {Pan},
  \citenamefont {Xie}, \citenamefont {Wu},\ and\ \citenamefont
  {Sarma}}]{pan2022topological}%
  \BibitemOpen
  \bibfield  {author} {\bibinfo {author} {\bibfnamefont {H.}~\bibnamefont
  {Pan}}, \bibinfo {author} {\bibfnamefont {M.}~\bibnamefont {Xie}}, \bibinfo
  {author} {\bibfnamefont {F.}~\bibnamefont {Wu}},\ and\ \bibinfo {author}
  {\bibfnamefont {S.~D.}\ \bibnamefont {Sarma}},\ }\bibfield  {title} {\bibinfo
  {title} {Topological phases in ab-stacked {MoTe2/WSe2}: {Z2} topological
  insulators, {Chern} insulators, and topological charge density waves},\
  }\href@noop {} {\bibfield  {journal} {\bibinfo  {journal} {Phys. Rev. Lett.}\
  }\textbf {\bibinfo {volume} {129}},\ \bibinfo {pages} {056804} (\bibinfo
  {year} {2022})}\BibitemShut {NoStop}%
\bibitem [{\citenamefont {Xie}\ \emph {et~al.}(2022)\citenamefont {Xie},
  \citenamefont {Zhang}, \citenamefont {Hu}, \citenamefont {Mak},\ and\
  \citenamefont {Law}}]{xie2022valley}%
  \BibitemOpen
  \bibfield  {author} {\bibinfo {author} {\bibfnamefont {Y.-M.}\ \bibnamefont
  {Xie}}, \bibinfo {author} {\bibfnamefont {C.-P.}\ \bibnamefont {Zhang}},
  \bibinfo {author} {\bibfnamefont {J.-X.}\ \bibnamefont {Hu}}, \bibinfo
  {author} {\bibfnamefont {K.~F.}\ \bibnamefont {Mak}},\ and\ \bibinfo {author}
  {\bibfnamefont {K.~T.}\ \bibnamefont {Law}},\ }\bibfield  {title} {\bibinfo
  {title} {Valley-polarized quantum anomalous hall state in moir{\'e}
  {MoTe2/WSe2} heterobilayers},\ }\href@noop {} {\bibfield  {journal} {\bibinfo
   {journal} {Phys. Rev. Lett.}\ }\textbf {\bibinfo {volume} {128}},\ \bibinfo
  {pages} {026402} (\bibinfo {year} {2022})}\BibitemShut {NoStop}%
\bibitem [{\citenamefont {Pan}\ \emph {et~al.}(2020)\citenamefont {Pan},
  \citenamefont {Wu},\ and\ \citenamefont {Sarma}}]{pan2020band}%
  \BibitemOpen
  \bibfield  {author} {\bibinfo {author} {\bibfnamefont {H.}~\bibnamefont
  {Pan}}, \bibinfo {author} {\bibfnamefont {F.}~\bibnamefont {Wu}},\ and\
  \bibinfo {author} {\bibfnamefont {S.~D.}\ \bibnamefont {Sarma}},\ }\bibfield
  {title} {\bibinfo {title} {{Band topology, Hubbard model, Heisenberg model,
  and Dzyaloshinskii-Moriya interaction in twisted bilayer WSe2}},\ }\href@noop
  {} {\bibfield  {journal} {\bibinfo  {journal} {Phys. Rev. Res.}\ }\textbf
  {\bibinfo {volume} {2}},\ \bibinfo {pages} {033087} (\bibinfo {year}
  {2020})}\BibitemShut {NoStop}%
\bibitem [{\citenamefont {Angeli}\ and\ \citenamefont
  {MacDonald}(2021)}]{angeli2021gamma}%
  \BibitemOpen
  \bibfield  {author} {\bibinfo {author} {\bibfnamefont {M.}~\bibnamefont
  {Angeli}}\ and\ \bibinfo {author} {\bibfnamefont {A.~H.}\ \bibnamefont
  {MacDonald}},\ }\bibfield  {title} {\bibinfo {title} {$\gamma$ valley
  transition metal dichalcogenide moir{\'e} bands},\ }\href@noop {} {\bibfield
  {journal} {\bibinfo  {journal} {Proc. Natl. Acad. Sci. U.S.A.}\ }\textbf
  {\bibinfo {volume} {118}},\ \bibinfo {pages} {e2021826118} (\bibinfo {year}
  {2021})}\BibitemShut {NoStop}%
\bibitem [{\citenamefont {Xian}\ \emph {et~al.}(2021)\citenamefont {Xian},
  \citenamefont {Claassen}, \citenamefont {Kiese}, \citenamefont {Scherer},
  \citenamefont {Trebst}, \citenamefont {Kennes},\ and\ \citenamefont
  {Rubio}}]{xian2021realization}%
  \BibitemOpen
  \bibfield  {author} {\bibinfo {author} {\bibfnamefont {L.}~\bibnamefont
  {Xian}}, \bibinfo {author} {\bibfnamefont {M.}~\bibnamefont {Claassen}},
  \bibinfo {author} {\bibfnamefont {D.}~\bibnamefont {Kiese}}, \bibinfo
  {author} {\bibfnamefont {M.~M.}\ \bibnamefont {Scherer}}, \bibinfo {author}
  {\bibfnamefont {S.}~\bibnamefont {Trebst}}, \bibinfo {author} {\bibfnamefont
  {D.~M.}\ \bibnamefont {Kennes}},\ and\ \bibinfo {author} {\bibfnamefont
  {A.}~\bibnamefont {Rubio}},\ }\bibfield  {title} {\bibinfo {title}
  {Realization of nearly dispersionless bands with strong orbital anisotropy
  from destructive interference in twisted bilayer {MoS2}},\ }\href@noop {}
  {\bibfield  {journal} {\bibinfo  {journal} {Nat. Commun.}\ }\textbf {\bibinfo
  {volume} {12}},\ \bibinfo {pages} {5644} (\bibinfo {year}
  {2021})}\BibitemShut {NoStop}%
\bibitem [{\citenamefont {Anderson}\ \emph {et~al.}(2023)\citenamefont
  {Anderson}, \citenamefont {Fan}, \citenamefont {Cai}, \citenamefont
  {Holtzmann}, \citenamefont {Taniguchi}, \citenamefont {Watanabe},
  \citenamefont {Xiao}, \citenamefont {Yao},\ and\ \citenamefont
  {Xu}}]{anderson2023programming}%
  \BibitemOpen
  \bibfield  {author} {\bibinfo {author} {\bibfnamefont {E.}~\bibnamefont
  {Anderson}}, \bibinfo {author} {\bibfnamefont {F.-R.}\ \bibnamefont {Fan}},
  \bibinfo {author} {\bibfnamefont {J.}~\bibnamefont {Cai}}, \bibinfo {author}
  {\bibfnamefont {W.}~\bibnamefont {Holtzmann}}, \bibinfo {author}
  {\bibfnamefont {T.}~\bibnamefont {Taniguchi}}, \bibinfo {author}
  {\bibfnamefont {K.}~\bibnamefont {Watanabe}}, \bibinfo {author}
  {\bibfnamefont {D.}~\bibnamefont {Xiao}}, \bibinfo {author} {\bibfnamefont
  {W.}~\bibnamefont {Yao}},\ and\ \bibinfo {author} {\bibfnamefont
  {X.}~\bibnamefont {Xu}},\ }\bibfield  {title} {\bibinfo {title} {Programming
  correlated magnetic states with gate-controlled moir{\'e} geometry},\
  }\href@noop {} {\bibfield  {journal} {\bibinfo  {journal} {Science}\ ,\
  \bibinfo {pages} {eadg4268}} (\bibinfo {year} {2023})}\BibitemShut {NoStop}%
\bibitem [{\citenamefont {Cai}\ \emph {et~al.}(2023)\citenamefont {Cai},
  \citenamefont {Anderson}, \citenamefont {Wang}, \citenamefont {Zhang},
  \citenamefont {Liu}, \citenamefont {Holtzmann}, \citenamefont {Zhang},
  \citenamefont {Fan}, \citenamefont {Taniguchi}, \citenamefont {Watanabe}
  \emph {et~al.}}]{cai2023signatures}%
  \BibitemOpen
  \bibfield  {author} {\bibinfo {author} {\bibfnamefont {J.}~\bibnamefont
  {Cai}}, \bibinfo {author} {\bibfnamefont {E.}~\bibnamefont {Anderson}},
  \bibinfo {author} {\bibfnamefont {C.}~\bibnamefont {Wang}}, \bibinfo {author}
  {\bibfnamefont {X.}~\bibnamefont {Zhang}}, \bibinfo {author} {\bibfnamefont
  {X.}~\bibnamefont {Liu}}, \bibinfo {author} {\bibfnamefont {W.}~\bibnamefont
  {Holtzmann}}, \bibinfo {author} {\bibfnamefont {Y.}~\bibnamefont {Zhang}},
  \bibinfo {author} {\bibfnamefont {F.}~\bibnamefont {Fan}}, \bibinfo {author}
  {\bibfnamefont {T.}~\bibnamefont {Taniguchi}}, \bibinfo {author}
  {\bibfnamefont {K.}~\bibnamefont {Watanabe}}, \emph {et~al.},\ }\bibfield
  {title} {\bibinfo {title} {Signatures of fractional quantum anomalous {Hall}
  states in twisted {MoTe2}},\ }\href@noop {} {\bibfield  {journal} {\bibinfo
  {journal} {Nature}\ ,\ \bibinfo {pages} {1}} (\bibinfo {year}
  {2023})}\BibitemShut {NoStop}%
\bibitem [{\citenamefont {Foutty}\ \emph {et~al.}(2023)\citenamefont {Foutty},
  \citenamefont {Kometter}, \citenamefont {Devakul}, \citenamefont {Reddy},
  \citenamefont {Watanabe}, \citenamefont {Taniguchi}, \citenamefont {Fu},\
  and\ \citenamefont {Feldman}}]{foutty2023mapping}%
  \BibitemOpen
  \bibfield  {author} {\bibinfo {author} {\bibfnamefont {B.~A.}\ \bibnamefont
  {Foutty}}, \bibinfo {author} {\bibfnamefont {C.~R.}\ \bibnamefont
  {Kometter}}, \bibinfo {author} {\bibfnamefont {T.}~\bibnamefont {Devakul}},
  \bibinfo {author} {\bibfnamefont {A.~P.}\ \bibnamefont {Reddy}}, \bibinfo
  {author} {\bibfnamefont {K.}~\bibnamefont {Watanabe}}, \bibinfo {author}
  {\bibfnamefont {T.}~\bibnamefont {Taniguchi}}, \bibinfo {author}
  {\bibfnamefont {L.}~\bibnamefont {Fu}},\ and\ \bibinfo {author}
  {\bibfnamefont {B.~E.}\ \bibnamefont {Feldman}},\ }\bibfield  {title}
  {\bibinfo {title} {Mapping twist-tuned multi-band topology in bilayer
  {WSe2}},\ }\href@noop {} {\bibfield  {journal} {\bibinfo  {journal}
  {arXiv:2304.09808}\ } (\bibinfo {year} {2023})}\BibitemShut {NoStop}%
\bibitem [{\citenamefont {Yu}\ \emph {et~al.}(2020)\citenamefont {Yu},
  \citenamefont {Chen},\ and\ \citenamefont {Yao}}]{yu2020giant}%
  \BibitemOpen
  \bibfield  {author} {\bibinfo {author} {\bibfnamefont {H.}~\bibnamefont
  {Yu}}, \bibinfo {author} {\bibfnamefont {M.}~\bibnamefont {Chen}},\ and\
  \bibinfo {author} {\bibfnamefont {W.}~\bibnamefont {Yao}},\ }\bibfield
  {title} {\bibinfo {title} {Giant magnetic field from moir{\'e} induced
  {Berry} phase in homobilayer semiconductors},\ }\href@noop {} {\bibfield
  {journal} {\bibinfo  {journal} {Natl. Sci. Rev.}\ }\textbf {\bibinfo {volume}
  {7}},\ \bibinfo {pages} {12} (\bibinfo {year} {2020})}\BibitemShut {NoStop}%
\bibitem [{\citenamefont {Haavisto}\ \emph {et~al.}(2022)\citenamefont
  {Haavisto}, \citenamefont {Lado},\ and\ \citenamefont
  {Otero~Fumega}}]{haavisto2022topological}%
  \BibitemOpen
  \bibfield  {author} {\bibinfo {author} {\bibfnamefont {M.}~\bibnamefont
  {Haavisto}}, \bibinfo {author} {\bibfnamefont {J.~L.}\ \bibnamefont {Lado}},\
  and\ \bibinfo {author} {\bibfnamefont {A.}~\bibnamefont {Otero~Fumega}},\
  }\bibfield  {title} {\bibinfo {title} {Topological multiferroic order in
  twisted transition metal dichalcogenide bilayers},\ }\href@noop {} {\bibfield
   {journal} {\bibinfo  {journal} {SciPost Phys.}\ }\textbf {\bibinfo {volume}
  {13}},\ \bibinfo {pages} {052} (\bibinfo {year} {2022})}\BibitemShut
  {NoStop}%
\bibitem [{\citenamefont {Zeng}\ \emph {et~al.}(2023)\citenamefont {Zeng},
  \citenamefont {Xia}, \citenamefont {Kang}, \citenamefont {Zhu}, \citenamefont
  {Kn{\"u}ppel}, \citenamefont {Vaswani}, \citenamefont {Watanabe},
  \citenamefont {Taniguchi}, \citenamefont {Mak},\ and\ \citenamefont
  {Shan}}]{zeng2023integer}%
  \BibitemOpen
  \bibfield  {author} {\bibinfo {author} {\bibfnamefont {Y.}~\bibnamefont
  {Zeng}}, \bibinfo {author} {\bibfnamefont {Z.}~\bibnamefont {Xia}}, \bibinfo
  {author} {\bibfnamefont {K.}~\bibnamefont {Kang}}, \bibinfo {author}
  {\bibfnamefont {J.}~\bibnamefont {Zhu}}, \bibinfo {author} {\bibfnamefont
  {P.}~\bibnamefont {Kn{\"u}ppel}}, \bibinfo {author} {\bibfnamefont
  {C.}~\bibnamefont {Vaswani}}, \bibinfo {author} {\bibfnamefont
  {K.}~\bibnamefont {Watanabe}}, \bibinfo {author} {\bibfnamefont
  {T.}~\bibnamefont {Taniguchi}}, \bibinfo {author} {\bibfnamefont {K.~F.}\
  \bibnamefont {Mak}},\ and\ \bibinfo {author} {\bibfnamefont {J.}~\bibnamefont
  {Shan}},\ }\bibfield  {title} {\bibinfo {title} {Integer and fractional
  {Chern} insulators in twisted bilayer {MoTe2}},\ }\href@noop {} {\bibfield
  {journal} {\bibinfo  {journal} {arXiv:2305.00973}\ } (\bibinfo {year}
  {2023})}\BibitemShut {NoStop}%
\bibitem [{\citenamefont {Park}\ \emph {et~al.}(2023)\citenamefont {Park},
  \citenamefont {Cai}, \citenamefont {Anderson}, \citenamefont {Zhang},
  \citenamefont {Zhu}, \citenamefont {Liu}, \citenamefont {Wang}, \citenamefont
  {Holtzmann}, \citenamefont {Hu}, \citenamefont {Liu}, \citenamefont
  {Taniguchi}, \citenamefont {Watanabe}, \citenamefont {haw Chu}, \citenamefont
  {Cao}, \citenamefont {Fu}, \citenamefont {Yao}, \citenamefont {Chang},
  \citenamefont {Cobden}, \citenamefont {Xiao},\ and\ \citenamefont
  {Xu}}]{park2023observation}%
  \BibitemOpen
  \bibfield  {author} {\bibinfo {author} {\bibfnamefont {H.}~\bibnamefont
  {Park}}, \bibinfo {author} {\bibfnamefont {J.}~\bibnamefont {Cai}}, \bibinfo
  {author} {\bibfnamefont {E.}~\bibnamefont {Anderson}}, \bibinfo {author}
  {\bibfnamefont {Y.}~\bibnamefont {Zhang}}, \bibinfo {author} {\bibfnamefont
  {J.}~\bibnamefont {Zhu}}, \bibinfo {author} {\bibfnamefont {X.}~\bibnamefont
  {Liu}}, \bibinfo {author} {\bibfnamefont {C.}~\bibnamefont {Wang}}, \bibinfo
  {author} {\bibfnamefont {W.}~\bibnamefont {Holtzmann}}, \bibinfo {author}
  {\bibfnamefont {C.}~\bibnamefont {Hu}}, \bibinfo {author} {\bibfnamefont
  {Z.}~\bibnamefont {Liu}}, \bibinfo {author} {\bibfnamefont {T.}~\bibnamefont
  {Taniguchi}}, \bibinfo {author} {\bibfnamefont {K.}~\bibnamefont {Watanabe}},
  \bibinfo {author} {\bibfnamefont {J.}~\bibnamefont {haw Chu}}, \bibinfo
  {author} {\bibfnamefont {T.}~\bibnamefont {Cao}}, \bibinfo {author}
  {\bibfnamefont {L.}~\bibnamefont {Fu}}, \bibinfo {author} {\bibfnamefont
  {W.}~\bibnamefont {Yao}}, \bibinfo {author} {\bibfnamefont {C.-Z.}\
  \bibnamefont {Chang}}, \bibinfo {author} {\bibfnamefont {D.}~\bibnamefont
  {Cobden}}, \bibinfo {author} {\bibfnamefont {D.}~\bibnamefont {Xiao}},\ and\
  \bibinfo {author} {\bibfnamefont {X.}~\bibnamefont {Xu}},\ }\bibfield
  {title} {\bibinfo {title} {Observation of fractionally quantized anomalous
  {Hall} effect},\ }\href@noop {} {\bibfield  {journal} {\bibinfo  {journal}
  {arXiv:2308.02657}\ } (\bibinfo {year} {2023})}\BibitemShut {NoStop}%
\bibitem [{\citenamefont {Li}\ \emph {et~al.}(2021{\natexlab{d}})\citenamefont
  {Li}, \citenamefont {Kumar}, \citenamefont {Sun},\ and\ \citenamefont
  {Lin}}]{li2021spontaneous}%
  \BibitemOpen
  \bibfield  {author} {\bibinfo {author} {\bibfnamefont {H.}~\bibnamefont
  {Li}}, \bibinfo {author} {\bibfnamefont {U.}~\bibnamefont {Kumar}}, \bibinfo
  {author} {\bibfnamefont {K.}~\bibnamefont {Sun}},\ and\ \bibinfo {author}
  {\bibfnamefont {S.-Z.}\ \bibnamefont {Lin}},\ }\bibfield  {title} {\bibinfo
  {title} {Spontaneous fractional chern insulators in transition metal
  dichalcogenide moir{\'e} superlattices},\ }\href@noop {} {\bibfield
  {journal} {\bibinfo  {journal} {Phys. Rev. Res.}\ }\textbf {\bibinfo {volume}
  {3}},\ \bibinfo {pages} {L032070} (\bibinfo {year}
  {2021}{\natexlab{d}})}\BibitemShut {NoStop}%
\bibitem [{\citenamefont {Wang}\ \emph {et~al.}(2023)\citenamefont {Wang},
  \citenamefont {Zhang}, \citenamefont {Liu}, \citenamefont {He}, \citenamefont
  {Xu}, \citenamefont {Ran}, \citenamefont {Cao},\ and\ \citenamefont
  {Xiao}}]{wang2023fractional}%
  \BibitemOpen
  \bibfield  {author} {\bibinfo {author} {\bibfnamefont {C.}~\bibnamefont
  {Wang}}, \bibinfo {author} {\bibfnamefont {X.-W.}\ \bibnamefont {Zhang}},
  \bibinfo {author} {\bibfnamefont {X.}~\bibnamefont {Liu}}, \bibinfo {author}
  {\bibfnamefont {Y.}~\bibnamefont {He}}, \bibinfo {author} {\bibfnamefont
  {X.}~\bibnamefont {Xu}}, \bibinfo {author} {\bibfnamefont {Y.}~\bibnamefont
  {Ran}}, \bibinfo {author} {\bibfnamefont {T.}~\bibnamefont {Cao}},\ and\
  \bibinfo {author} {\bibfnamefont {D.}~\bibnamefont {Xiao}},\ }\bibfield
  {title} {\bibinfo {title} {Fractional chern insulator in twisted bilayer
  {MoTe2}},\ }\href@noop {} {\bibfield  {journal} {\bibinfo  {journal}
  {arXiv:2304.11864}\ } (\bibinfo {year} {2023})}\BibitemShut {NoStop}%
\bibitem [{\citenamefont {Reddy}\ \emph {et~al.}(2023)\citenamefont {Reddy},
  \citenamefont {Alsallom}, \citenamefont {Zhang}, \citenamefont {Devakul},\
  and\ \citenamefont {Fu}}]{reddy2023fractional}%
  \BibitemOpen
  \bibfield  {author} {\bibinfo {author} {\bibfnamefont {A.~P.}\ \bibnamefont
  {Reddy}}, \bibinfo {author} {\bibfnamefont {F.~F.}\ \bibnamefont {Alsallom}},
  \bibinfo {author} {\bibfnamefont {Y.}~\bibnamefont {Zhang}}, \bibinfo
  {author} {\bibfnamefont {T.}~\bibnamefont {Devakul}},\ and\ \bibinfo {author}
  {\bibfnamefont {L.}~\bibnamefont {Fu}},\ }\bibfield  {title} {\bibinfo
  {title} {Fractional quantum anomalous {Hall} states in twisted bilayer
  {MoTe2} and {WSe2}},\ }\href@noop {} {\bibfield  {journal} {\bibinfo
  {journal} {arXiv:2304.12261}\ } (\bibinfo {year} {2023})}\BibitemShut
  {NoStop}%
\bibitem [{\citenamefont {Jiang}\ \emph {et~al.}(2018)\citenamefont {Jiang},
  \citenamefont {Zhou}, \citenamefont {Dai},\ and\ \citenamefont
  {Wang}}]{jiang2018antiferromagnetic}%
  \BibitemOpen
  \bibfield  {author} {\bibinfo {author} {\bibfnamefont {K.}~\bibnamefont
  {Jiang}}, \bibinfo {author} {\bibfnamefont {S.}~\bibnamefont {Zhou}},
  \bibinfo {author} {\bibfnamefont {X.}~\bibnamefont {Dai}},\ and\ \bibinfo
  {author} {\bibfnamefont {Z.}~\bibnamefont {Wang}},\ }\bibfield  {title}
  {\bibinfo {title} {Antiferromagnetic {Chern} insulators in noncentrosymmetric
  systems},\ }\href@noop {} {\bibfield  {journal} {\bibinfo  {journal} {Phys.
  Rev. Lett.}\ }\textbf {\bibinfo {volume} {120}},\ \bibinfo {pages} {157205}
  (\bibinfo {year} {2018})}\BibitemShut {NoStop}%
\bibitem [{\citenamefont {Xiao}\ \emph {et~al.}(2012)\citenamefont {Xiao},
  \citenamefont {Liu}, \citenamefont {Feng}, \citenamefont {Xu},\ and\
  \citenamefont {Yao}}]{xiao2012coupled}%
  \BibitemOpen
  \bibfield  {author} {\bibinfo {author} {\bibfnamefont {D.}~\bibnamefont
  {Xiao}}, \bibinfo {author} {\bibfnamefont {G.-B.}\ \bibnamefont {Liu}},
  \bibinfo {author} {\bibfnamefont {W.}~\bibnamefont {Feng}}, \bibinfo {author}
  {\bibfnamefont {X.}~\bibnamefont {Xu}},\ and\ \bibinfo {author}
  {\bibfnamefont {W.}~\bibnamefont {Yao}},\ }\bibfield  {title} {\bibinfo
  {title} {Coupled spin and valley physics in monolayers of {MoS2} and other
  {group-VI} dichalcogenides},\ }\href@noop {} {\bibfield  {journal} {\bibinfo
  {journal} {Phys. Rev. Lett.}\ }\textbf {\bibinfo {volume} {108}},\ \bibinfo
  {pages} {196802} (\bibinfo {year} {2012})}\BibitemShut {NoStop}%
\bibitem [{\citenamefont {Chatterjee}\ \emph {et~al.}(2020)\citenamefont
  {Chatterjee}, \citenamefont {Bultinck},\ and\ \citenamefont
  {Zaletel}}]{chatterjee2020symmetry}%
  \BibitemOpen
  \bibfield  {author} {\bibinfo {author} {\bibfnamefont {S.}~\bibnamefont
  {Chatterjee}}, \bibinfo {author} {\bibfnamefont {N.}~\bibnamefont
  {Bultinck}},\ and\ \bibinfo {author} {\bibfnamefont {M.~P.}\ \bibnamefont
  {Zaletel}},\ }\bibfield  {title} {\bibinfo {title} {Symmetry breaking and
  skyrmionic transport in twisted bilayer graphene},\ }\href@noop {} {\bibfield
   {journal} {\bibinfo  {journal} {Phys. Rev. B}\ }\textbf {\bibinfo {volume}
  {101}},\ \bibinfo {pages} {165141} (\bibinfo {year} {2020})}\BibitemShut
  {NoStop}%
\bibitem [{\citenamefont {Kane}\ and\ \citenamefont
  {Mele}(2005{\natexlab{a}})}]{kane2005quantum}%
  \BibitemOpen
  \bibfield  {author} {\bibinfo {author} {\bibfnamefont {C.~L.}\ \bibnamefont
  {Kane}}\ and\ \bibinfo {author} {\bibfnamefont {E.~J.}\ \bibnamefont
  {Mele}},\ }\bibfield  {title} {\bibinfo {title} {Quantum spin {Hall} effect
  in graphene},\ }\href@noop {} {\bibfield  {journal} {\bibinfo  {journal}
  {Phys. Rev. Lett.}\ }\textbf {\bibinfo {volume} {95}},\ \bibinfo {pages}
  {226801} (\bibinfo {year} {2005}{\natexlab{a}})}\BibitemShut {NoStop}%
\bibitem [{\citenamefont {Kane}\ and\ \citenamefont
  {Mele}(2005{\natexlab{b}})}]{kane2005z}%
  \BibitemOpen
  \bibfield  {author} {\bibinfo {author} {\bibfnamefont {C.~L.}\ \bibnamefont
  {Kane}}\ and\ \bibinfo {author} {\bibfnamefont {E.~J.}\ \bibnamefont
  {Mele}},\ }\bibfield  {title} {\bibinfo {title} {{Z2} topological order and
  the quantum spin hall effect},\ }\href@noop {} {\bibfield  {journal}
  {\bibinfo  {journal} {Phys. Rev. Lett.}\ }\textbf {\bibinfo {volume} {95}},\
  \bibinfo {pages} {146802} (\bibinfo {year} {2005}{\natexlab{b}})}\BibitemShut
  {NoStop}%
\bibitem [{\citenamefont {Hohenadler}\ \emph {et~al.}(2011)\citenamefont
  {Hohenadler}, \citenamefont {Lang},\ and\ \citenamefont
  {Assaad}}]{hohenadler2011correlation}%
  \BibitemOpen
  \bibfield  {author} {\bibinfo {author} {\bibfnamefont {M.}~\bibnamefont
  {Hohenadler}}, \bibinfo {author} {\bibfnamefont {T.}~\bibnamefont {Lang}},\
  and\ \bibinfo {author} {\bibfnamefont {F.}~\bibnamefont {Assaad}},\
  }\bibfield  {title} {\bibinfo {title} {Correlation effects in quantum
  spin-hall insulators: A quantum monte carlo study},\ }\href@noop {}
  {\bibfield  {journal} {\bibinfo  {journal} {Phys. Rev. Lett.}\ }\textbf
  {\bibinfo {volume} {106}},\ \bibinfo {pages} {100403} (\bibinfo {year}
  {2011})}\BibitemShut {NoStop}%
\bibitem [{\citenamefont {Hohenadler}\ \emph {et~al.}(2012)\citenamefont
  {Hohenadler}, \citenamefont {Meng}, \citenamefont {Lang}, \citenamefont
  {Wessel}, \citenamefont {Muramatsu},\ and\ \citenamefont
  {Assaad}}]{hohenadler2012quantum}%
  \BibitemOpen
  \bibfield  {author} {\bibinfo {author} {\bibfnamefont {M.}~\bibnamefont
  {Hohenadler}}, \bibinfo {author} {\bibfnamefont {Z.}~\bibnamefont {Meng}},
  \bibinfo {author} {\bibfnamefont {T.}~\bibnamefont {Lang}}, \bibinfo {author}
  {\bibfnamefont {S.}~\bibnamefont {Wessel}}, \bibinfo {author} {\bibfnamefont
  {A.}~\bibnamefont {Muramatsu}},\ and\ \bibinfo {author} {\bibfnamefont
  {F.}~\bibnamefont {Assaad}},\ }\bibfield  {title} {\bibinfo {title} {Quantum
  phase transitions in the {Kane-Mele-Hubbard} model},\ }\href@noop {}
  {\bibfield  {journal} {\bibinfo  {journal} {Phys. Rev. B}\ }\textbf {\bibinfo
  {volume} {85}},\ \bibinfo {pages} {115132} (\bibinfo {year}
  {2012})}\BibitemShut {NoStop}%
\bibitem [{\citenamefont {Rachel}\ and\ \citenamefont
  {Le~Hur}(2010)}]{rachel2010topological}%
  \BibitemOpen
  \bibfield  {author} {\bibinfo {author} {\bibfnamefont {S.}~\bibnamefont
  {Rachel}}\ and\ \bibinfo {author} {\bibfnamefont {K.}~\bibnamefont
  {Le~Hur}},\ }\bibfield  {title} {\bibinfo {title} {Topological insulators and
  {Mott} physics from the {Hubbard} interaction},\ }\href@noop {} {\bibfield
  {journal} {\bibinfo  {journal} {Phys. Rev. B}\ }\textbf {\bibinfo {volume}
  {82}},\ \bibinfo {pages} {075106} (\bibinfo {year} {2010})}\BibitemShut
  {NoStop}%
\bibitem [{\citenamefont {Zare}\ and\ \citenamefont
  {Mosadeq}(2021)}]{zare2021spin}%
  \BibitemOpen
  \bibfield  {author} {\bibinfo {author} {\bibfnamefont {M.-H.}\ \bibnamefont
  {Zare}}\ and\ \bibinfo {author} {\bibfnamefont {H.}~\bibnamefont {Mosadeq}},\
  }\bibfield  {title} {\bibinfo {title} {Spin liquid in twisted homobilayers of
  {group-VI} dichalcogenides},\ }\href@noop {} {\bibfield  {journal} {\bibinfo
  {journal} {Phys. Rev. B}\ }\textbf {\bibinfo {volume} {104}},\ \bibinfo
  {pages} {115154} (\bibinfo {year} {2021})}\BibitemShut {NoStop}%
\bibitem [{Note1()}]{Note1}%
  \BibitemOpen
  \bibinfo {note} {The argument can be easily generalized to summations of band
  energies of nearly degenerate bands}\BibitemShut {NoStop}%
\bibitem [{sup()}]{supp}%
  \BibitemOpen
  \href@noop {} {}\bibinfo {note} {Supplemental Material containing an
  additional phase diagram for the Kane-Mele-Hubbard model.}\BibitemShut
  {Stop}%
\end{thebibliography}%

\end{document}